# Whose AI Dream? In search of the aspiration in data annotation.

Ding Wang
Google Research
Bangalore, India
drdw@google.com

Shantanu Prabhat
Google Research
Bangalore, India
shprabhat@google.com

Nithya Sambasivan
Google Research
Bangalore, India
nithyasamba@google.com

## ABSTRACT

Data is fundamental to AI/ML models. This paper investigates the work practices concerning data annotation as performed in the industry, in India. Previous human-centred investigations have largely focused on annotators' subjectivity, bias and efficiency. We present a wider perspective of the data annotation: following a grounded approach, we conducted three sets of interviews with 25 annotators, 10 industry experts and 12 ML/AI practitioners. Our results show that the work of annotators is dictated by the interests, priorities and values of others above their station. More than technical, we contend that data annotation is a systematic exercise of power through organizational structure and practice. We propose a set of implications for how we can cultivate and encourage better practice to balance the tension between the need for high quality data at low cost and the annotators' aspiration for well-being, career perspective, and active participation in building the AI dream.

## CCS CONCEPTS

• **Human-centered computing** → **Empirical studies in HCI**; **Field studies**.

## KEYWORDS

data annotation, AI labour, future of work, qualitative study

## 1 INTRODUCTION

Artificial Intelligence (AI), Machine Learning (ML), and Internet of Things (IoT) have already been on a fast-growing trajectory, and the COVID-19 pandemic has further accelerated the pace of adoption of automation in healthcare, education, entertainment, and more [9, 16]. Machine Learning (ML), the most popular subset of AI techniques, depends on data to 'learn', that is, to uncover patterns, make classifications or predict future outcomes. Therefore, data, particularly the labeled and annotated datasets, are fundamental to the development and success of ML models.

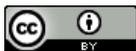

Data annotation and labeling work, predominantly done manually, has historically been dependent on and carried out by independent workers on crowd-work platforms like Amazon Mechanical Turk, Appen, or Clickworker [28, 37]. However, there has been a rise in private *annotation firms* dedicated to providing data labelling and annotation services. These are third-party companies, who employ full-time staff for annotation work and take up contracts for annotation and labeling tasks in bulk. Annotation firms employ thousands of people as data annotators [33, 55]. Annotation firms scale to global customers, serving technology clients like Microsoft, TripAdvisor, and eBay. The third-party data labeling solutions market is projected to grow to 4.1 billion USD by 2024 [51]. The rapid rise of the data annotation market is not embraced by the industry alone. Governments from the 'Global South' have taken notice of the demand and are investing in data annotation as the welfare and economic growth programs. For instance, China has established dozens of government-endorsed poverty alleviation programs partnering with Chinese tech companies (e.g., Alibaba and JD) to bring data labeling jobs to remote and rural parts [58]. At the same time, India's national AI strategy report praises the importance of data labeling in generating employment [35]. Data annotation is now a mainstream part of the AI/ML economy, owing in part to the rise of annotation firms.

The work of data annotation has considerably different characteristics at a firm compared to platform-based labeling work. First, annotation firms employ full-time workforces, in contrast to the freelancing model of platform-based work. Annotation firms have payrolls with fixed salaries, provide computerized equipment, and enjoy well-organized work structures and benefits. In contrast, platform-based workers have been reported to be under-paid and exploited in precarious work conditions with no employee benefits [22, 28]. Second, the initial capital required for a third-party annotation company is considerably smaller, thus more accessible for entrepreneurs to set up such start-ups. However, platform-based labeling requires heavy resources for setup and to mediate the match between clients with data needs and workers with the right skills. Finally, annotation companies provide ancillary services such as project management for the annotation tasks, dedicated client communication channels and touch-pints, quality control mechanisms, and privacy risk mitigation (such as NDAs), etc. [1] This balance between the quality and cost that the annotation companies promise attracts more data requesters to choose them over platforms. In contrast, platforms can only offer labeling or annotation services.

Compared to the well-celebrated market value of data annotation, the human labor required to sift through and sort data and ultimately train these systems remains under-recognized (beyond a few notable exceptions (e.g., [14, 28, 31, 46]). The annotation industry employs thousands of workers worldwide to read texts, view images and video, and label data to enable the models that AI systems rely upon. A straightforward example here is the work involved in labeling faces in photographs. Although many computer vision systems tout the incredible performance of algorithms that identify faces, attention is rarely given to work involved in labeling the data—despite its importance in training and refining the models. Complicating such labor processes are the norms imposed on labelers, in the form of labeling tasks and processes, which are often homogeneous and standardized across labelers, e.g., in the normalization of image classification. The enforcement of homogeneity of norms ignores frictions between multiple layers of meaning and, often, culturally-sensitive value systems [39, 50]. Prior HCI and CSCW scholarship has examined the experiences of platform-based workers [28, 31]. However, we know little about how annotation work structures, work practices, and personal experiences of data annotators employed as full-time office workers. How do the material, infrastructural, and organizational factors of data annotation firms affect the worker?

---

[1] There are also annotation companies that adopt an impact sourcing model, which aims to provide job opportunities to the population historically excluded from formal employment (e.g., marginalized communities in rural areas and people with disabilities) and prioritize worker training and developing local expertise.

In this paper, we investigate the work practices concerning data annotation as performed in the industrial context of India. We report from in-depth, semi-structured interviews with three stakeholder groups of data annotation: twenty-five data annotators (in India), ten managers coordinating annotation requests (global, including India, Kenya, Nepal, Philippines, Nigeria, and the US), and twelve AI/ML engineers (in the US, Israel, and the UK) whose work relied on annotated data sets. [2] We examine the work practices and experiences starting from recruitment of the workers, to the orientation and on-job training, to their evaluation and career progression, juxtaposed with the operational insights and understanding of the industry landscape from the managers as well as the requesters' needs and priorities when procuring data work. Our research reveals that despite the transition from platform-facilitated, task-based piece work to full-time employment, the working conditions for individual annotators remained problematic. For instance, the pursuit of high quality in their work output translated into layers of scrutiny and created immense work pressure. Annotator jobs were still subjected to great precarity. There was little career progression for individual annotators to either continue in annotation or transition out to technical positions. The rapid pace of growth of the data annotation industry did not translate to benefit the individual annotators, as it did other stakeholders. For instance, the average tenure for an annotator contract was between 12-18 months, which offered little stability and opportunities for long-term career development. Annotators were often engineering graduates who imagined data annotation as a gateway to working on cutting-edge AI applications, such as self-driving cars. Annotators reported facing enormous work pressure to meet unrealistic deadlines, laid out by those above their station, to prioritize clients' interests and timelines. A common practice shared by annotation companies was the '*ever-increasing target rate*': as soon as the annotators were able to complete the daily tasks, the target for completion they had to meet increased.

Despite the professionalization of data annotation jobs, the pathways for growth are broken, workers are still anxious about employment and performance. There are admittedly advantages such professionalization has over the platform. The worker benefit comes with formal employment (e.g., access to pension and insurance). Overall, the lack of professional growth, tightened control through organizational structure, and the unpaid and much-expected overwork paint a grim picture. We discuss implications for the organizations, operations, and data set tasks in working with annotators. We intend this paper to make the following contributions to HCI and CSCW:

- It provides a detailed empirical investigation that identifies characteristics of being an annotator, conducted with full-time employees, fixed workplaces, and well-organized working structures. Our work further motivates a critical mode of inquiry into the organization of work and labor in data production and how they spread or evolve into other forms across the gig economy.
- An understanding that professionalizing data annotation jobs can positively or negatively impact various stakeholders such as individual annotators, data requesters, and annotation companies.
- We propose high-level recommendations for engaging with organized data annotation, particularly its organization, operation, and relevance to the growing importance of ethical practices in data.

## 2 RELATED WORK
### 2.1 The Work of Data

The remarkable capacities of AI-infused and data-intensive systems are regularly lauded not only by the digerati but in the popular press. Far less recognized, beyond some notable exceptions (e.g. [14, 28, 31, 46]), is the human labor required to sift through and sort data and ultimately train these systems. Data work is under-appreciated at a structural level in AI [49]. Crowd-sourcing platforms worldwide employ thousands of workers to read texts, view images and video, and label data to produce the models that AI systems rely on. Without the work humans pour into making sense and labeling the data to train and refine it, even the most advanced systems could not make a prediction [40, 41, 44]. Complicating such labors are the norms that are imposed on annotators. This enforcement in the homogeneity of norms used, for example, in the normalization of image classification mask frictions between multiple layers of meaning and, often, culturally sensitive value systems [20, 38, 39, 50]. To define such a 'norm' itself even with in an organization is no easy task. Muller et al. presented a detail account of the challenges faced and mitigation techniques adopted by the data science teams while defining ground-truth from an organizational perspective [41]. That the process to reach the desirable ground-truth is far from straight-forward and it requires rounds of iteration , improvisation, and collaboration and coordination beyond the labeling team [ibid].

The invisible work and hidden labor in crowd-sourced micro-tasks have been well discussed in HCI/CSCW [28, 29, 31, 37]. Though the visibility of work is complex, as de Carvalho argued, it is the result of both the politics of the work and the content itself [17]. As Suchman argues the underlying work practices often get obscured by the representation of work and the assumptions about how work tasks are performed by various actors other than the worker themselves [54]. Similarly, Barley and Kunda suggest that how we talk about work "indexes our assumptions about how work is organized" [1]. 'When work occurs in a technological platform that splits or masks workers' is one of the conditions Poster et al. argue that would make work invisible [13]. They further argue that the lack of transparency is often accelerated by the platforms that 'hide the nature of the labor process itself, including the location of the work; the conditions under which it is done; and the ethnicity, race, and nationalities of the workers' [ibid]. Our study continues this discussion on the visibility of work by adding an empirical corpus from the annotators whose work is also made hidden. This provides the indispensable context and sensibility to the context [5, 21] that in which more detailed design for annotation and annotators take place.

---
[2] In this paper, we primarily focus on the annotators' perspectives and the interviews with the managers and requesters are used to supplement the annotators' stories.

## 2.2 Changing Trends

Past work in HCI/CSCW and beyond has drawn attention to precisely these hidden and often exploitative labors [2, 28, 31, 46]. Our work investigates a contemporary moment where the global gig economy relies on new organizational actors emerging in the Global South [42]. Smaller platforms, such as iMerit in India [3], CloudFactory in Nepal [4], Samasource in Kenya [5], are creating new models to compete with the likes of Amazon's Mechanical Turk. Based on a very different business model, these start-ups moved away from sourcing and providing microtasks for their labelers and towards an overtly ethical model that prioritizes worker training and developing local expertise.

Meanwhile, we have also been witnessing the change of traditional work landscape – a trend towards 'taskification' of work, that instead of full-time workers, companies prefer project- and task-oriented contractual hiring, and the availability of necessary digital infrastructure to match clients with freelancers online efficiently and at scale [17, 24, 28]. When work is task-oriented, the individual's expertise is valued over past organizational and educational credentials [23]. The emphasis on individual expertise is echoed in Blaising et al.'s work uncovering the strategies adopted by freelancers to manage long-term career trajectories [3].

As the historically task-based, crowd-sourced, platform-facilitated work becomes more organized and structured [42, 57] the traditional work is being taskified [8, 18, 23, 24]. In our study, we saw both trends intertwined. While the organizations participating in hiring workers through structured employment to perform previously crowd-sourced tasks on the increase, the workers were guaranteed stability through the employment necessarily and were anticipated to become part of the 'crowd' eventually.

---

[3]https://imerit.net/
[4]https://www.cloudfactory.com/
[5]https://www.sama.com/

## 2.3 Towards ethical and responsible AI

Previous academic work has provided a critical discussion on the politics involved in data-driven systems [15, 19, 36]. What's highlighted is also the discussion on investigating the capitalistic logics embedded into them [4, 10]. However, such discussion has to go beyond the data and systems in question. The critical perspective ought to be extended to the process of the work involved in the production of data that powers these systems to avoid data cascades of downstream harm [49]. Although, in light of more discussion on ethical and responsible AI, data documentation such as datasheet [25] and data nutrition label [30], efforts to document the sociological aspects of data [32] and documentation to promote reflexivity [39] still put the emphasis on data, and its impact on model design and development, rarely does the practice and process during the data annotation or the data workers take the center of the stage. A recent discussion on AI Ethics detailed the concrete steps we ought to follow but still does not include the discussion on data practice [53]. What is encouraging is that the discourse on ethical AI practices has moved beyond a mere discussion. In an unprecedented draft of Internet Information Service Algorithm Recommendation Management Regulations published by China's cyberspace watchdog, the Cyberspace Administration of China (CAC) [7], the 17th article specifically commented on the worker and labor involved in AI recommendation service.

> The provider of algorithmic recommendation service who provides work scheduling services to workers shall establish and improve related algorithms such as platform order distribution, remuneration composition, payment, working hours, rewards and punishments, and fulfill the obligations of protecting workers' rights and interests.

Though not explicitly spelled out, the workers referred to here are gig workers dependent on the platform economy. Annotators can be part of this new regulative protection depending on their employment capacity. Our research argues that practices adopted in data annotation, beyond the annotation itself, from recruitment to training to evaluation, should also appear in the discussion around ethical and responsible AI as they are an integral part of how the data comes into being. Without the work and labor that were poured into the data annotation process, ML efforts are no more than sandcastles.

## 3 METHODOLOGY

This paper defines data annotation as the sense-making practice of a given dataset, where annotators assign meaning to data using (pre-defined) labels. Previous human-centered investigations have primarily focused on annotators' subjectivities, biases, and efficiency. We propose a broader and more holistic view of the data annotation practice: guided by a grounded approach [26, 27], we conducted a qualitative study consisting of three sets of in-depth and semi-structured with twenty-five annotators from India, ten industry experts who operate and manage data annotation companies across the globe (e.g., India, Kenya, Nepal, Philippines, and the US) and twelve experts who are engineers (majority ML/AI practitioners) whose work relies on annotated data sets from India, the US, and Israel during December 2020 to July 2021.

We collaborated with two third-party research recruitment agencies on the recruitment for all three different types of participants. In addition, we worked with one recruitment agency based in India to recruit the annotators. Unlike the studies on crowd workers or freelancers, where the participant recruitment advertisement could be posted as a job on the platforms where the crowd workers and freelancers find jobs, it was proven to be difficult to break into the annotators' circle (especially during the pandemic). Our sampling strategy included four main criteria:

- to include annotators from diverse areas of expertise and skill;
- to include both annotators who are relatively new to the field (having experience of three months to one year) and those who have been a part of it for several years;
- to include annotators from different levels of seniority and work cycle, i.e., from the ones who work on annotation, to quality control to team management.

However, despite the desire to cover a diverse area of expertise and skills, most of our annotator participants specialized in image/video annotation for the automotive industry due to the time constraint and access. We chose India because the country is home to one of the largest annotation labor markets in the world [47]. The third-party recruitment agency helped liaise and arrange interviews with the interested participants who expressed their willingness to participate in the study. Our sample consisted of fourteen male and eleven female participants across India (see Table 1 in the next section for details). Typical tasks they had done thus far during their annotation career included drawing bounding boxes, identifying semantic and polygonal segmentation, annotating general images and video, labeling entities, and categorizing content and text for different projects and clients, and industries. The majority of our participants worked to support the autonomous vehicle industry, with only a few working in what could be categorized as advertising and marketing. Given the geographical spread of our participants and the raging pandemic at the time in India, interviews were conducted online using video conferencing software. Interviews were scheduled based on participants' convenience and conducted in English. Informed written consent was electronically obtained from all interviewees via the third-party agency prior to the commencement of interviews. Separate consent was obtained for recording the interviews. The audiotapes were transcribed verbatim subsequently. The consent process involved explicitly informing freelancers that their participation, responses, and duration of engagement were entirely voluntary. At the outset, participants were informed that they could stop at any time or refuse to answer any questions. Although the interviews ranged from 35 to 70 minutes, this did not affect their compensation rate, nor were they rated differently. All annotator interviewees received a flat rate of 1500 INR (roughly 20 USD) as compensation for participation. This amount was in escrow, and participants knew beforehand that they would get paid regardless of how long they engaged with us. Therefore, the research team took these steps to diffuse potential power imbalances as per standard practices. The interview protocol covered a range of topics such as their work/job, their motivations to take up annotation jobs, the challenges they experienced with it, their use of annotation tools and software, what they liked about the work and what they did not, their experiences working with different kinds of companies, clients and tasks, how they managed their work, and how they were managed, recruited, and trained.

Interview transcripts were then analyzed collaboratively by all authors to identify relevant themes. The analysis was consistent with and inspired by the ethnomethodological ethnographies in HCI [11, 12, 43]. Our analysis took a broadly ethnomethodological perspective. Ethnomethodologically-informed, ethnographies explicate the knowledgeable, artful ways in which workers orient to their work and how technologies and other artifacts are used as part of the methodical accomplishment of that work [6, 45]. As well as analyzing interview transcripts, we took the additional step to examine the tools used by the annotators through two walk-through sessions from the industry experts to get a holistic picture of the annotator's work. Ethnomethodological analyses of work are useful in generating a granular understanding of what activities constitute 'work' in a setting, how they are accomplished in practice, who is involved in this accomplishment, what resources are drawn upon, and what skills and tools are involved in mobilizing those resources [ibid]. Through this close look at the seemingly ordinary details, our analysis seeks to unveil not just what the world looks like but how it comes to look as it does. The emphasis is, in other words, on the detail of work as understood and interpreted by the people who perform it.

The data were analyzed by the first author individually and by the first author and the second author in analysis sessions explicating a particular topic, as is typical of the ethnomethodological approach. Since we adopted the 'grounded approach' [26, 27], the techniques of constant comparison and constant iteration (i.e., iterations of coding and re-coding) were used in the development of themes so as to avoid the classic problems of 'cumulation' and 'theoretical imperialism' – "an analytically imposed reconstruction of the procedures of a setting, insufficiently sensitive to the understandings of a setting's participants."[26] These analytic sessions allowed interesting topics to be identified and endogenous themes to emerge from the data (such as the annotation process, an annotator's aspiration, and their frustration). To stay true to the grounded approach, we were extremely cautious not to impose categories external to the data to codify the data. Ethnomethodological ethnographies are valuable in informing design [45], and we used the resulting understanding of the annotation work from all three perspectives to inspire a set of implications that aim to address some of the challenges annotators and requesters face and therefore to steer the discussion around the work annotation towards a more constructive and worker-centered direction.

## 4  PARTICIPANT BACKGROUNDS

We present in the following tables the detailed backgrounds of the three different set of participants that contributed to our study: the data annotators (in Table 1), the industry experts (in Table 2) and the ML/AI engineers whose work depend on annotated data (in Table 3).

## 5  FINDINGS

Our findings present different patterns and practices regarding the entire annotation process and seeing annotation as an industry. It is essential to point out that we look at the annotation as its industry consisting of a set of practices widely adopted as standards by different stakeholders in the industry. Through juxtaposing the themes that emerged from the set of interviews with the annotators to the ones that came from the study with the data requesters and industry experts,[6] we demonstrate the widely accepted yet questionable industry practices, such as the hiring practice and quality control; and the potential catalysts for these practices, e.g., the pursuit of near perfection in data quality and the desire for diversity.

| P# | Employment Status | Highest Qualification | Role | Experience |
| --- | --- | --- | --- | --- |
| P1 | Freelancer | BE (EC) | Data Annotator | 1 year 4 months |
| P2 | Internship | BE (EC) | Data Annotator | 1 month |
| P3 | Employed- Full time | BE (Mechanical) | Data Quality Analyst | 3 years 3 months |
| P4 | Employed- Full time | B Tech | Data annotator | 4 years |
| P5 | Seeking for a job | B Tech (CS) | Engineer data operations | 3 years |
| P6 | Seeking for a job | BE | Annotator and validator | 3 years |
| P7 | Employed- Full time | B E (CS) | Senior Data annotator | 1 year 6 months |
| P8 | Employed- Full time | B E (CS) | Senior data annotator | 1 year 2 months |
| P9 | Seeking for a job | Masters (CS) | Senior executive | 2 years 7 months |
| P10 | Employed- Full time | BE (EC) | Data annotator | 1 year |
| P11 | Seeking for a job | BE (CS) | Process executive | 1 year 6 months |
| P12 | Employed- Full time | BE (EC) | Data annotation Engineer | 1 year 2 months |
| P13 | Employed- Full time | BE | Process executive | 3 years 3 months |
| P14 | Employed- Full time | BE (EC) | Data Engineer | 3 years 9 months |
| P15 | Employed- part time | BE | Annotation engineer | 1 year 2 months |
| P16 | Employed- part time | Pursuing MBA | Anotation Lead | 5 years 5 months |
| P17 | Employed- part time | BE | Annotation engineer | 4 years 5 months |
| P18 | Employed- Full time | BE | Data Engineer | 2 years 3 months |
| P19 | Employed- Full time | BE | Quality Analyst | 2 years |
| P20 | Seeking for a job | BE (Civil) | Data annotation Engineer | 1 year 9 months |
| P21 | Employed- Full time | B Tech | Trainee Graphic Designer | 1 year 2 months |
| P22 | Seeking for a job | BE (EC) | Senior associate | 2 years 4 months |
| P23 | Employed- Full time | BE (Mechanical) | Data Engineer | 2 years 8 months |
| P24 | Employed- Full time | BE (CS) | Data Annotation Engineer | 1 year 2 months |
| P25 | Employed- Full time | BE (Mechanical) | Data Annotation Engineer | 1 year 6 months |

Table 1: Annotator Participant's Qualification, Role and Experience

| P# | Position/Role | Country | Area |
| --- | --- | --- | --- |
| E1 | Senior Associate Manager of Operations | Kenya | Data Annotation |
| E2 | Program Manager for ML research | India | Mobile Application |
| E3 | Global Director | Nepal | Data Annotation |
| E4 | Product Manager (manages Data Annotation needs) | US | Machine Learning |
| E5 | Head of Customer Recruit | Kenya | Data Annotation |
| E6 | Senior Customer Success Engineer | India | Data Annotation |
| E7 | Custom Recruit/President and CEO | Philippines | Data Annotation |
| E8 | Adjunct Professor (manages Data Annotation Projects for research) | US | Machine Learning |
| E9 | Chief Operation Manager | Nigeria | Data Annotation |
| E10 | Co-founder (manages Data Annotation Needs) | US | Machine Learning |

Table 2: Industry Experts who operates and manages data annotation projects

---

[6] In this paper, we put a conscious emphasis on the annotators' stories. We present the industry experts' perspective to supplement and contrast the annotators' account (sections 5.1 to 5.3). In 5.4, we highlight a more distinct view from the data requesters to highlight the competing priorities that led to the status quo of the annotation practice and process.

| P# | Role | Country | Nature of Work | Annotation Type |
|---|---|---|---|---|
| R1 | Programme Manager | US | ML Product Development | Text, Image |
| R2 | Research Scientist | Israel | Research | Text |
| R3 | Research Scientist | UK | Research | Text, Audio |
| R4 | Product Manager | Israel | Research | GIS Data, Image |
| R5 | Consultant | US | ML Product Development | Image |
| R6 | Technical Lead | US | ML Product Development | Text, Image |
| R7 | Technical Lead | US | ML Product Development | Image, Text |
| R8 | Product Manager | US | ML Product Development | Text |
| R9 | Professor/Entrepreneur | US | Research/ML Product Development | Image, Text, Audio |
| R10 | Head of NLP | UK | ML Product Development | Text |
| R11 | Data Scientist | US | ML Product Development | Text, Image |
| R12 | CEO | US | ML Product Development | Text |

Table 3: Practitioner in AI/ML whose work depends on annotated datasets

## 5.1 Becoming an annotator – a gateway to AI?

*5.1.1 The recruitment.* The first thing that stood out about our annotator participants was their qualifications. All twenty-five annotators we interviewed had undergraduate degrees in B Tech or BE (i.e., technology and engineering), several majored in computer science, one even had a Master's degree, and one at the time of the interview was pursuing an MBA degree. Although Ross et al. [48] showed that the crowd-workers from developing countries are more likely to have a Bachelor's degree, we were surprised by how well (if not over) qualified the annotators are. We picked this pattern from the recruitment stage and probed our participants during the interview to understand whether this was a selection bias – that the recruitment agency we collaborated with was only looking for participants with Bachelor's degrees. As it turned out, this was not a bias introduced during the participant recruitment process but a status quo of the annotation industry in India— having a Bachelor's degree is a mere entry requirement to become an annotator with the annotation companies. Interestingly, the emphasis on a higher educational background is valued over having relevant previous experiences. When asked what the requirement was to become an annotator, whether the relevant previous experience was either necessary or desirable, all of our participants reported no previous experience required to get their first annotation jobs but having a bachelor's degree in technology or engineering is a must. Having such a high entry bar in regards to educational background is not unique to the annotation industry in India. Our interviews with the industry experts revealed that this is a common practice among annotation companies in Nepal, Kenya, and the Philippines too, as one of our experts put it

> Most of our hires would have an undergraduate degree. We hire them directly from the campus, some might be still doing their degree, but that is okay. Bottom-line is, they will have to have a high school diploma at the very least and good, good at what they do. (Expert 2)

This is echoed by Expert 1, who provided us with a potential explanation for insisting on requiring undergraduate degrees. For the company, this requirement serves as a short-cut to effectively filter out candidates that do not meet the literacy requirements – reading and writing literacy in English to comprehend and complete tasks, computer literacy to operate the devices and tools for work, and the digital and technical literacy to grasp and contextualize what their tasks mean to the systems dependent on the annotated data. What appears to be a 'short-cut' for the annotation companies significantly narrowed the access to the employment generated by annotation. The access is not defined by the ability of the potential worker but by the convenience and interest of the annotation firms. Additionally, at times, having previous experience in annotation was seen as a disadvantage because experienced annotators would expect higher pay. As one of our annotators explained her struggle to get employed as an experienced annotator:

> oh, they won't prefer the experienced people, they prefer for freshers itself, so that they can pay less for fresher. Experienced people will also get hired but it depends. Here they took experienced [annotators] but only for two months. Like the project is only for two months, so they hired experience so that they can give how much they want, but for long project I don't [think] they hire experienced. (P15)

However, the fact that the job as an annotator required no previous experience was appreciated by the first-timer annotators.

> Like it's easy compared to other technologies and all and it is like there's no risk factor in this and like I like to do is job. It's easy to work without any stress and all. No need of any other knowledge much knowledge. Just the basic thing you need like computer knowledge and little bit of the … what the client needs based on that we work. (P4)

Since it is not required for the prospect annotators to have previous experience, many took annotations as their first employment after graduation. From the interviews, we learned that usually, there are three common paths for a potential job seeker to come across annotation jobs – campus recruitment organized by the annotation companies each year, online job search, and internal referral. Out of the three paths, the referral was the most common way for our annotator participants to find their employments at the time.

Nearly half of them were referred to their companies by a friend, a classmate or alumni. What's particularly interesting was that when asked about the motivation behind joining the annotation industry, almost all of our participants [7] expressed their excitement and desire to be part of this new industry that is part of the AI boom happening in India and this is also a gateway for them to be connected to the global AI market. As one of our participant said

> I heard like this was the upcoming booming technology. It was new technology like in India for data annotation thing. It was like just a new thing. So I thought like it may be like, it has some growth in annotation. That is why I joined this. A few company has annotations. (P1)

What made this AI boom more convincing and attractive was that rapid growth was not just seen in flashy marketing material. It was witnessed and experienced by our participants. As P19 noted, the number of annotation companies in the little town on the west coast of India, where he lived, grew from 2-3 when he started to 6-8 after only a year. To him (or to any), this was the sign of a growing field. Another noted the quick growth of the company he works for as the project intake increased and project portfolio widened,

> It depends, means a project, like first they only started with a traffic light marking. Later, they got the project from the same client as pedestrian one marking then, then sign marking and some computer graphics like some other computer graphics project also there, and so the amount of data coming is very very large. So only few people can complete that data in a particular time, so they increase the head count. Some hundred were added.

The newness and the promise of a bright future that comes with the shiny new tech industry attracted our participants to become annotators. However, this narrative around AI, annotation, and technology (autonomous vehicle in particular) at large is purposely crafted and marketed to the annotators. More than half the annotators we interviewed were drawn initially to annotation because of its close connection to AI and ML, that it is fundamental to building the future of 'driver-less cars chauffeuring people around'. P10 recalled that during his interview, he was asked what autonomous driving cars he liked, in addition to the questions about how tagged images can be used to train the models autonomous cars depend on. As this advertisement depicts, a typical annotation job post portrays the job as a well-paid, reputable, and noble part of this more significant industry, a great company, and a fancy AI dream (Figure 1).

Generally, there were multiple rounds of interviews. First, there were interviews where the prospect annotators were introduced to annotation, and some were interviewed to perform example tasks to test their ability to interpret the annotation requirements and execute them. What is particularly interesting is that, in addition to the more related to annotation, there is a technical round where the prospect annotators are interviewed for their technical capabilities i.e., writing codes in various programming languages (e.g., C, Java, and Python etc). Some participants were made to write elementary

---

[7] apart from the one who was working at an annotation intern who also happened to be the only one working in the annotation for digital marketing at the time

**BIG DATA OF AI BUSINESS**

**₹15K - ₹30K** a month is the starting salary of a data labeller

**$150 million** is what the data labelling market was worth in 2018

**$1 billion** is what research company Cognilytica estimates it to be by the end of 2023

**Figure 1: A typical annotation job advertisement in India [33]**

C programs (check for even or odd). Others were hired because they listed relevant skills like "CAD" on the job sites. Though our initial thought was that it might be a practice that a few companies in the industry adopted, to our surprise that this was not at all a unique practice to some annotation companies. Instead, this was a widespread phenomenon across all the participants and their companies. The technical round to test their coding ability during the interview has been normalized as industry-standard. Some participants explained that some open-sourced annotation interfaces might require coding ability to have additional extensions for the tasks at hand, though what is required to program these extensions is HTML or Python, rather than C or Java. We learned that none of our participants ever needed to code or program during their tenure as annotators. Many have mentioned that the technical teams would handle the software or interface design and development. However, by having a technical round, the annotator job appeared instantly to be more technical, and the selection process became more complex and rigorous at the same time.

*5.1.2 The training process.* Given that most of the annotators went into the field without previous experience or even knowledge about what annotation entailed, it made training an essential step for them to transition into being an annotator. We found two types of training for the annotators, orientation training for when they first joined a company and pre-project training for whenever they were assigned to a new project. Almost all the companies that the annotators worked for provided mandatory orientation training. The majority of the orientation training usually takes two weeks to complete, though there are two companies that the annotators worked for that provided longer training which lasted three to four weeks. During the orientation training, the emphasis was placed on getting the annotators familiarised with the tools, tasks, and processes. Depending on the company, the orientation training takes different forms. Some companies organized the orientation training as a presentation given by the team lead, some companies used pre-recorded videos, and some solely relied on the annotators to go through the training brochures on their own. According to the annotators, these training materials usually consist of samples of what data annotation requirements look like, a step-by-step guide on completing an annotation task (e.g., drawing bounding boxes on a pedestrian in an image), and how to interact and operate the tools for annotation (e.g., how to upload the data to the system, how to use the interface and how to export the file). In addition to the more theoretical training, the annotators also need to take on practical training. Again, it is the learning process; the annotators will go through a data set demo to apply what they have learned from the training material in practice. The orientation training aimed to ramp the annotators up to the pace to handle actual annotation work. It clearly emphasized the practical use of the tool and how to get the work done.

> Training is like first time they show the raw data and they've these how to how to use the tool because the tool have lots of, uhm, loads of attributes and, uhm, means functions, uhm, like tool every time the tool is getting updated like that is the one thing. Then they also teach how to tag the object. I know what we need to concentrate on annotation time and how to start the annotation and yeah like that. (P19)

Although the emphasis on the connection between annotation and AI was central to the narrative to get the annotators on board, during the orientation training, there was little mention of either the AI or its dependency on the data the annotators were to work on. As our participant noted,

> They just give me the brief introduction on how the machine works and how the deep learning works. They give you more information on how the image segmentation [an annotation technique] works then image like... how to split frames [in a video]... and a little bit knowledge of machine learning. (P3)

The orientation training covered the general knowledge required to complete the tasks, and due to the diversity of annotation tasks and the speed at which annotation tasks, techniques, and tools evolve, the training material is often out of date by the time it was delivered to the annotators. Hence, it was not a surprise that some found it redundant,

> like, it's easy and no need of training and just, just required detailed manual instructions only that is necessary and the tools which they want us to use. (P1)

Detailed instruction for the tasks at hand provides more value than the training materials. The skills required for data annotation for different projects were ubiquitous, and once the annotators had experience with one project, they could do another project.

> Actually, the only thing we want is the detailed eye, and interested in the tool and the techniques. if we actually annotation, how they using some three four techniques like bounding box an notation, semantic annotation and cuboid annotation polygon annotation. For means they are marking like a traffic signs traffic signals pedestrians. So for each project either issues bounding box annotation or semantic annotation. So if we worked in one project we can also do another project also. (P19)

As the quote demonstrates, though the training can be helpful, the implicit and tacit knowledge they take from one project to another as they gather more experience. The annotator then gave a specific example of such tacit knowledge.

> I found that... if we get an image [folder]. We like tag first image then next second image. Actually. That is the wrong way, first thing you notice the entire images, entire images in that folder and if we get one image, we have to check means we have to we have to analyze that image like from my point of view start from left to right or right to left then only we can reduce the mistakes. (P19)

He summarised through experience that the specific procedure guarantees better output and quality. This is tacit knowledge essential to the annotators' work yet not included in the orientation training. Another annotator shared proudly that it took him two weeks to figure out, on his own, all the keyboard short-cuts in the software his company used to perform annotation tasks. This discovery dramatically improved his productivity. Though crucial to one's productivity, the software short-cuts were not part of the orientation training.

The pre-project training before each project was where this kind of tacit knowledge could often get passed from one annotator to another as it was often led by the team lead or project lead at a team level. The purpose of this training was to get the annotators instructed with the specific dataset and requirements for the project they were about to work on. This training was relatively short; it usually takes a few days to get all the annotators on the project up to speed with the annotation guidelines and the data. Companies sometimes skip the pre-project training to compress the project time to impress the client (this was particularly true to the companies that praised themselves for turnaround speed), especially when the project was under a time crunch and extra workers were dropped halfway in order to speed up the delivery process.

> That was not like training thing. Just we used to sit with the one who was working there also the new joiners . But they also had the same training and we used to absorb everything. (P15)

> Joining mid project means no proper training, the training was done through sitting next to the experience annotators on the project and observe. (P22)

Though training is essential to the annotators' work, under the enormous delivery pressure, the step often gets omitted to save time. However, as we learned from the annotators, skipping the training can be counterproductive. As the knowledge transfer from the organization to the worker and among workers regarding procures, protocols and tactics are the cornerstones of annotation quality and productivity. Compared to the client's request, the annotators' interest remained secondary to the annotation companies. The data quality is regarded as a set of measurable performance metrics, i.e., accuracy rate.

### 5.2 Being an annotator

*5.2.1 Day in the life for an annotator.* An annotator usually spent 8 to 10 hours (sometimes 12 hours) daily working on projects. However, through the interview, we learned that the working hour is not the hour spent in the office but rather the hours spent on tasks, which indicates the actual working hour was likely more protracted than what the annotators reported.

> It was 9:00 [am] to 6:00 [pm] before [in the previous company], but here [in the current one] also the timings has a 9 to 6 itself. But somewhat work pressure is the know like need to deliver the project early. So it takes 10 hours, 11 hours. (P15)

When we asked whether the annotators were compensated for extra working hours, they all said no.

All except two of our annotator participants worked on-site in their company offices or on their client's premises prior to the pandemic lock-down. By the time of our interview, almost all who had previously worked in an office who still had an annotation job had returned to their offices (this was before the deadly second wave hit India). Their day typically started with a status check led by the team or project lead, either as a meeting or through an email. Firstly, they reviewed any pending tasks from the day before needed rework or completion, then the lead handed out the data of the day waiting to be labeled to the annotators and walked through the update related to the tools and attributes. After which, the annotators would begin their day of work on the data sets to meet the daily target set by the company. The annotators first familiarized themselves with the data set and the annotation guideline. Then they began annotating each image, a frame of a video, or a snippet of text according to the instruction, e.g., identifying the object or keyword and assigning attributes to them. Before the day ended, a quality analyst would collect all the data from the annotators and check for quality. The errors and feedback (i.e., issues that were not mistakes but poor practices) highlighted during the quality check will be sent back to the annotators as pending tasks to complete the first thing the next day and incorporate in future work.

Apart from the only freelancing annotator and the annotator intern, every annotator we interviewed worked with a laptop provided by their employers. These laptops were usually under close control by the employer apart from the websites and soft-wares deemed relevant to their work, such as Microsoft Teams, Google chat, and WhatsApp in sporadic cases. These communication tools are vital to the annotators as they were the primary venue where they usually raise quires regarding the task, the tool they use, and other work-related issues. Even though most companies prohibit personal communication tools such as WhatsApp and Facebook Messenger for work, it is common for annotators to create WhatsApp group chats for work.

> We are connected on WhatsApp and we have a group, we all the, all the labelers and our seniors for are there, [through] which we are connected and if we get any difficulty facing you regarding the server, regarding the labeling or anything. If any issue is there then our seniors are there and they are very sweet and they totally help us. (P2)

The access to unauthorized websites or soft-wares was blocked, making it nearly impossible for them to use their laptop outside work or working hours for other purposes. The annotators often had experience with a wide variety of annotations tools, ranging from open-source tools (e.g., CVAT, Data Turks and Labellmg, etc.) to in-house tools developed by the annotation company to client tools. From both the interviews with the annotators and the industry experts, we realized that more data annotation companies started developing their annotation tools (some with data analytics features) to provide data annotation as an end-to-end service.

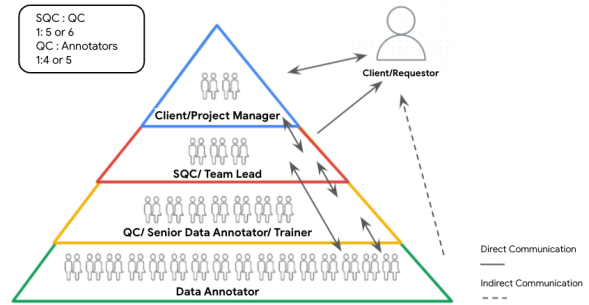

**Figure 2: Organisational Hierarchy In Annotation in Indian Companies**

*5.2.2 More than technical.* The process of data annotation, during which the annotators work to assign meanings to the images, texts, and audios, might be technical. However, we contend that data annotation is more than technical, that it is a systematic exercise of organizational structure and power (Figure 2). This is consistent with what Miceli et al. [38] have observed in the data annotation companies in Argentina and Bulgaria.

As mentioned in the above section that the annotators had to meet a target (often daily and in some companies weekly) in terms of both the quantity (i.e., target rate) and the quality (i.e., accuracy rate) of tasks. The daily target rate is not a randomly assigned number. Often it was the team lead based on the past project experience to decide the daily target rate considering the turnaround time requested by the client. Another way to set the target rate, in general, was to find an average task completion rate among annotators. There were two popular ways of target setting. The first way was to find the best-performing, average-performing, and worst-performing annotators (based on the past performance on speed and accuracy) to work through a random sample from a new set of data. The target rate would be the average quantity of the tasks completed by all three types of annotators in a set time (an hour or a day).The other way was simply sending the sample data out for annotation to a team of annotators and taking that average. Given the average completion rate being set as the target, while there were annotators who exceeded the target, there were annotators who would inevitably struggle to meet this goal. Either way, the annotators did not get a say in a reasonable daily target rate. To make the matter worse, a common practice among the data annotation companies was to increase the target as soon as most of the annotators were able to hit the rate.

> Per day, it was initially not fixed and often that one week 30, and again, the next week is 40 again 50. 65 was the last. 65 images was target, but for QC it was 250 images. (15)

As our participant noted, as the average annotators chased the ever-increasing target rate, the rate for a quality analyst was stable though considerably larger. It is worth pointing out that the rejected tasks from the day before that required reworking were out of the daily target count. Meeting this target was a norm among annotators and taken for granted by their companies. For the annotators, it was not a matter ever up for negotiation. Many were taken aback when asked what would happen if someone did not make the target. As P15 explained, "Nothing, like they don't say anything but it will be assigned for us now, so we need to complete." As for the excellent annotators who consistently meet their target, their efforts would be acknowledged in appraisals in meetings or emails; there were no other rewards or bonuses.

Delivering high-quality data at the lowest cost possible is at the core of many annotation companies. [8] Being able to ensure the annotation quality is key to annotation companies proactively. Wang et al. [57] also noted the companies' proactive monitoring of work quality. Organizationally, there were three quality control lead levels: quality control analysts (QC), senior quality control analysts (SQC), and the client or project manager. QCs were required to check every single attribution in every file from the annotators. Generally, in one annotation team, the ratio of QC to annotator was 1:4 or 5. After the QC completed the first round of checks, SQC stepped in (at some companies, the team lead played this role) to double-check the quality by random sampling the annotated and checked data set. Finally, the project manager performed the last level of quality control before 'shipping the data'. Usually, after these three rounds of quality control, the accuracy rate before the data shipping was around 98%. Valentine et al. [56] suggested that establishing organizational structures improved the task process and productivity. Some annotators (P19, P23 and P14) reported that their companies pushed this even further to zero error rates.

> If you are making some mistakes then higher people don't accept, it affects their mood. (P23)

> Like yeah, it depends on clients also for accuracy thing. Like if they expect more no we have to give. (P14)

Technique-wise, there were also automated checks in addition to the manual quality check. Both the annotators and experts noted that the right combination of manual and automated checks could drastically improve the accuracy rate and productivity. For instance, using a script to check for missing bounding boxes or attributes automatically and wrong attributes allowed the QCs and SQCs to manually check for, say, the size of a bounding box, which is more of a best practice issue than an accuracy issue. Additionally, these layers of quality checks, especially the ones done by humans, were essential to catch the human errors from the annotators and vital to catch errors caused by the annotation tools. For example, P 19 noted that the annotation tool would shift the bounding boxed out the place before it finally crashed. This was an error that the script could not catch, but a QC who knew about the occasional crashes would keep an eye on it.

At times, the annotators reported issues they encountered working with the client tool or their in-house development team. Though the SQC or the team lead often mediated this, the annotator reported up the chain following the hierarchy. Issues with the tool they worked with were common. They ranged from random crashes to missing attributes [9]; from the lack of support in the tool for the precision required to the format errors that'd occur during exporting annotated images. Two annotators commented that the tools from the 'first timer' clients would require many updates and subsequently more back and forth communication. Even though some tools included features for the annotators to raise the issue or share it with other annotators to search for a solution, the reporting and discussion regarding issues remained in the communication tools separate from the annotation tool. Screenshots were often used to provide context to the issues they raised. Only the issues that could not be resolved locally among the annotators, QCs, and team leads would reach the clients. The responses from the client-side were communicated down the pecking order, that "it will be explained from project Lead to the all the labelers" (P15).

*5.2.3  Too good for the job?* Our annotator participants shared a somewhat mixed view of annotation. Some, although critical, still held a positive overview of both their jobs and the industry, but for some, they became more disillusioned as they worked. The more disillusioned ones were disappointed mainly by the dissonance between the narrative of a fancy new industry and the often tedious day-to-day work. As some voiced,

> Anyone can do it, a 6th grader can do it... with some training, it's not difficult. Same thing over and over. (P8) No changes, each and every single day. (P7)

The repetitiveness and tedium of the job were noted by many. However, not all of their concern was on the monotony. Some were more worried that the lack of diversity in their portfolio would negatively impact their future employability.

> Technique is a technique, so if we learn more that will be good for our career... I didn't feel anything bore, but actually semantic segmentation is taking *much* time.

As it was not just Having exposure to a wide variety of annotation tasks, industries and techniques increased the likelihood for an annotator to go independent and their chance of getting hired by the next company. This promised one the 'sense of control' over their career, which is why P1 chose to become a freelancer. Annotators who worked on a diverse range of projects spoke fondly of the rotation between projects in some companies every a few months. The diversity in the projects provided the annotators more than the professional exposure they needed but sometimes a lens through which they were able to see the world. For example, P7 shared how much she enjoyed the video annotation project where she could see snow! However, working on a variety of projects is a luxury that was not the case for many.

The stress to meet the sometimes unrealistic target is another common cause for the annotators to feel dissatisfied.

> So much of work pressure. Like they wanted to deliver the project early. So it was quite a lot of work pressure. (P15)

---

[8]Quality and cost are the top two priority for ML practitioners, we present this in detail in section 5.4

[9]For instance, the attributes were required by the guideline yet not in the tool

Multiple annotators during the interview expressed their preference for 'easy tasks' that the tasks that took less time to complete were 'good tasks'. This preference was not due to laziness. Instead, it was a direct result of the often unrealistic target rate!

> So sometimes we can't reach the target, that is the issue in semantic segmentation... semantic segmentation is like pixel labeling, like we have to mark everything in that images like trees, roads sidewalk like everything, depends the project, but we have to tag it precisely. (P19)

Regardless of the clear difference between the two annotation techniques' efforts, the target image/frame rate was the same. The lack of clear instructions, low quality of the raw data, and the poor usability of the tools were the three leading causes for the annotators to miss their target on either quantity or quality that were also not entirely in their control. Their feedback rarely made it through the reporting chain and was addressed or implemented. P19 was the only annotator who shared a positive experience where his design suggestion to the UI of the client tool was adopted. He suggested adding a 'magnifying glass' icon to the UI, so it was more intuitive for the annotators to enlarge the otherwise small image they needed to annotate. More often than usual, the annotators' input was overlooked, especially when the annotation itself. P23 gave us an example when they were required to label the grass by the roadside. Upon submitting the annotated data, the client rejected the work because the dry grass was not separate from gravel, which was not part of the initial guideline. Depending on the image quality, it could be challenging to tell the dry grass from the gravel. The annotator explained that if the images were captured during the summer when the grass was drier, it could tell the difference, but their annotation had to cover all seasons. He was particularly frustrated when he discovered such fine detail would not have mattered in some cases. For example, for autonomous cars, there is a safe distance of 30cm between a detected object and the car. So when there was a visible fence in the image, there was no point in labeling grass from the gravel as the car would not be near it. This feedback was dismissed, and the motivation behind the feedback was regarded as pure laziness. These annotators were tested and hired for the technical ability to comprehend the AI and ML projects their data contributed to. However, ironically, it was not valued when they posed questions based on their understanding of ML and experience as annotators.

> If any object doesn't not belong to any class, there is a subclass called, no class, we tag that as no class actually. It is not... it doesn't means it's right. It's like the criteria from the client up to their liking. (P17)

For a machine to be able to see as a human, annotators have to see from the lens of the machine, make sense of the sight and label it accordingly. Unfortunately, the accuracy is often lost in this process that's not in the control of the annotators, yet they are the ones being blamed for the low accuracy and had to bear the consequence of it.

### 5.3 An annotator's aspiration

More than half of the annotators we interviewed saw data annotation as a preliminary job to becoming an engineer in AI and ML before entering the field. They soon understood the distance between annotation and ML. Despite the dependency AI and ML have on data annotation, at an individual level, the knowledge, skills, and experiences the annotators acquired during their tenure on data did not translate to what was required to become an ML engineer. This realization was what essentially made many of them stay on. We found that everyone, the annotators, experts, and requesters talked about the importance of 'having the right expertise'. However, there was no consensus on what qualified as expertise and what made the 'right'. The current articulation regarding an annotator's expertise was either a set of annotation techniques they mastered or the annotation tasks' domain. However, having experience annotating medical images does not make one a medical professional. The difficulty for them to break into the more technical side was real. Expert 1 estimated that roughly 10% of the annotators could take up more technical roles (e.g., software engineers and data analysts). However, we suspect the actual number was fewer. Only *one* out of the twenty-five annotators we interviewed made it into data science. This was not because of his experience as an annotator but rather the online courses he had taken on data analytics. When we asked our industry experts where the annotators would go, it was clear that the experts did not quite know. Though Expert 1 and 5 shared some anecdotal examples of previous annotators who were able to use the "capital" they accumulated working as an annotator to start their own small business (such as grocery shops), and some took the "experience and training" from the annotation companies and became freelancers. On the other hand, our annotators shared the other side of the story that was not as straightforward to go freelance. Experience-wise, lack of project diversity did not provide the competitive advantage they need, and the practical training in using annotation interfaces did not always transfer from one tool to another.

Annotation is a trade that indeed attested to 'practice makes perfect'. According to our annotators who had moved up the ladder once in a firm (22/25) and those above the annotator level (13/25), the promotion from annotators to QCs typically after six months, and one was promoted after only three months into the job. People quickly moved up the job ladder from annotators to QC and from QC to SQC. However, very few made the next promotion above SQC (2/25). The glass ceiling in annotation was low and fast to reach. If not moving up, then they would move around. The retention rate is not high. All eight industry experts who managed and operated data annotation companies confirmed that the annotators typically stay for 12-18 months. [10]. We contend that the current compensation system was not designed to keep annotators. P17 shared that her biggest frustration with her job was that there was no annual increment to her salary regardless of the companies she worked for. The salary increase happened only when they moved up the ladder. Thus, there was little encouragement to stay for the annotator who did not move the ladder. The aspiration to a normal career progression or stability in annotation was struck hard by the reality.

---

[10] Expert 6 provided a particular interesting note, that in his company, the in house annotators in their US offices had been there for more than three years and one even had tenure as old as the company which was five years. While the annotators through third party were always changing

> No, I don't want to work in this stream again. Like it is not safe... stable like, oh the project may end anytime and we will not be having the work. I like... it is not stable as other profiles. Like if they have the projects now they will be hiring lot of people's for that. Once the project has ended, they won't think there is no second thought of how continuing with the employees they were directly fire them off. (P15)

The pandemic added extra pressure. With 32 million Indians knocked out of the middle class in 2020, India contributed the largest share – about 60% – of the total global decline in the middle class during the pandemic [34]. As one of the worst-hit countries by COVID, this does not surprise. However, this is the reality that these annotators live in. Pursuing one's aspiration is pure luxury, as it requires additional financial and time investments while their livelihood was subject to changes often beyond their control.

> Due to this pandemic situation the company lost the project, and the company got shut down, then for two to three months I was not worked. (P22)

> Actually my first organization I lost my job in this this current situation after two months. I later I placed here. (P15)

At the time of our interview, about a third of our participants had experienced a job loss due to the pandemic, three were able to find another annotation job, and five were still job seeking.

Regardless of the hardship, most of our annotators shared a relatively positive sentiment towards annotation. They were very aware of their incremental contribution to AI and ML, and they took great pride in what they did.

> Before like we are not their data annotator, we can't... they cannot do something like machine learning. AI systems also not working without data so that same we are too proud of us, we are giving... that that much quality and that much productivity. P16

> Machine learning and AI system so we can't move without data. So what I am giving the data so machine learning and AI system are working. Without data they don't work. So what I'm saying, like I'm proud of to there like data labelling work. P23

> So it was an interesting thing and it was something like good to do because it is necessary. Like if some default is there then the car manufacturer will be like defect if based on the raw data is wrong. Then the machine will predict incorrect. So if it was like we need to do exactly what the given like zero, without zero percent error. (P1)

AI and ML promise to take away the repetitive, tedious, and dangerous work from humans, such as driving. However, it is currently at the cost of overqualified people performing tedious labor to build the AI dream.

### 5.4 From the requester's perspective

All twelve of the ML practitioners we interviewed ranked the quality of the data and the production cost associated with it as the top two things they cared about the most. They were also the most important criteria for these requesters to solicit a platform or a company to handle their data. There is a delicate balance between the requirement for high-quality data and the desire for a low cost that the requesters are fully aware of.

> It makes a difference where you go, you know, if you go to a [crowd-sourcing platform] which is something that we use extensively, it's good for certain things, it's cheap, it's not good for other things. And then, you know, we have to go to a different place. And we use [name of a third party annotation company] for that, which is much higher quality. At a cost though, of course. (R8)

The third-party annotation companies' main attraction is the right balance between quality and cost. Additionally, the requesters were also very conscious about the resources and time required to manage the data production. The annotation companies offered services beyond simply getting the data annotated compared to crowd-sourcing platforms. They provided the organizational structure to recruit and train the annotators, active monitoring and checks to ensure quality, and project management throughout the data production. To some requesters, the extra cost was worth it, as, after all, they are "engineers hired to build ML rather than managing the crowd workers!" In addition, the annotation companies helped them to address their previous concern with the lack of self-contained quality control mechanisms with the crowd-sourcing platforms.

The emphasis the requesters put on data quality manifested itself in rounds of quality checks at both the production and delivery end. The annotation companies tackled it with proactive monitoring reinforced through the organizational hierarchy during the production. The requesters created various technical solutions, the most common being scripts and plug-ins for automated quality control. R6, who provided their tool for data annotation, shared that their team would periodically embed random "hidden tests" among the ordinary tasks to get a sense of the annotators' average performance and catch "the ones that might be struggling". We have already learned from the industry experts that there were periodic exams on accuracy and speed to identify workers who might need extra training. However, the trust towards annotators from the requesters was not exceptionally high, and the fact that there was no direct channel between the annotators working the data and the requesters did not help. The data collection and data annotation process still mainly remained a black box for the requesters, i.e., they have little to no idea of who the annotators are, how they were trained, and how they carried out the tasks. Despite very few (2/12) regarded the interaction between the requesters and the taskers as unnecessary, more acknowledged that having more knowledge on the production of the data improves the overall quality (beyond just the one dataset but long-term collaboration). R6's company was the only one that actively shared the impact of the annotators' work with them, that R6 would update the annotators on the improvement of the model performance based on the data they worked hard on.

## 6 DISCUSSION

Data annotation, in our study, is the process of sense-making carried out by the annotators who are hired as full-time employees,

with fixed salaries, embedded in well-organised working structures. Third-party data annotation as an industry emerged at the back of the continuous expansion of AI and ML systems into more domains (such as an autonomous vehicle), and the growing investments in annotation from states only catalyzed it [35, 51, 58]. Our research indicates that despite the rapid growth in data annotation as an industry, its benefit didn't serve the individual annotators. It did other stakeholders. The experience, working as an annotator tells a different narrative to the well-celebrated success story of data annotation. Data annotation companies draw on the growth of historically crowd-sourced tasks to provide an alternative for AI/ML practitioners to find workers with the right set of skills. This alternative allows the practitioners to avoid decomposing complex tasks into smaller piece works, finding suitable solo workers with the corresponding skills, and project managing these tasks. In order to deliver such services, the data annotation companies set up a well-defined role and follow a clear organizational hierarchy that is tailored towards managing the complexities and high client expectations. Rigid organizational hierarchy does deliver more than desirable results to the requesters (e.g., higher data quality and faster turnaround). However, it impacts both the interpretation of the data [38] and the individual annotators. What is particularly problematic is that 'data quality' has been interpreted and measured as performance metrics i.e. accuracy rate. It speaks nothing to the skills the annotators bring or the perspectives they hold. Such a narrow interpretation of data quality is also detrimental to the dataset.

With the parameters of 'success' defined and decided, any scope for ingenuity is removed. Doing so eliminates nuances that might be beneficial to the dataset and the definition of quality. We found that our annotators rarely questioned the power that management and clients have over them in influencing their understanding of the data as well as their working conditions. Let us step through what this could mean in the specific situation that we have a very narrow idea on what it means to perform well that it is predetermined by the requesters in the Global North. There is little room for considerations and reflectections upon the misalignment between the annotators in the Global South and the requesters in the Global North. Additionally, none of them considered the technical tests during their interview unnecessary or unfair. They accepted the verbal appraisal and the promise of potential promotion as the appropriate rewards for excelling in their work.

We present implications for practitioners, researchers, institutions and organizations in supporting their aspirations while working with annotators. In line with what Dourish has spelled out in his seminal work, Implications for Design [21], the merit of this paper and the implications we propose are not on trivial recommendations for design. Rather, this paper attempts to provide the sensitivity of the context that designs for annotations are taking place. We echo Dourish's argument that the focus on implications for design is misplaced, as it misconstrues the nature of the ethnographic enterprise. Such misinterpretation of implication misses where ethnographic and qualitative inquiry can provide major insight and benefit for HCI research. As the value of ethnographic investigations for design goes beyond the specific design recommendations [5]. Ethnographic research does not rule out design options, but it points towards the consequences of the design choice that one makes [11]. It is to influence the design and development of computing systems by providing an in-depth understanding of how work gets done in practice in order to better support it. To sensitize designers and developers to the myriad issues that people who use technologies confront at both individual and organizational levels [ibid]. As we demonstrated, the annotation process presents a complex work setting. Before giving the annotators more tools, we need to understand the setting and the people, such as their aspirations and what they bring. [11]

## 6.1 From crowdwork to organised employment

The trend that the traditional employment is slowly being 'taskfied', full-time employment is on the decline while task-oriented contractual hiring is on the rise accelerated by the platforms facilitating the match between a job and a worker [17, 28]. From our study, we noticed a different, if not opposite, trend in the field of data annotation – instead of leaning more towards crowd work and microwork through platforms, the industry is showing a steady preference for organized labor, hiring annotators as full-time employees. Annotation firms offer stability and the usual benefits associated with employment (leaves, insurances, equipment, and office) for as long as the employment lasts. However, this trend does not eliminate the downsides that are associated with annotation work previously experienced by crowd workers, such as the precarity of the work (that the average tenure is between 12-18 months) and the undesirable working conditions (the long working hours, the unpaid extra labor, the constant scrutiny, and monitoring). Adding to the complexity, offering clear advantages over hiring solo-crowd workers with the add-on service values, the participation of third-party annotation companies in the traditional micro task space would inevitably make the market a more competitive place for solo workers. We found that a contributing factor of the short tenure is to keep the pool of annotators hired in one company constantly refreshed and updated. Though the annotation companies see it to achieve diversity, it is counterproductive. When the recruitment criteria and methods remain the same, a similar set of people would be added to the pool.

We found that as data annotation becomes organized employment, it didn't make the work annotators do visible. Rather it became more invisible. As we described in section 5.4, since the little interaction between a requester and an annotator that might happen before on a crowd-sourced platform being taken over by client managers in data annotation companies, the work that an annotator does become more opaque from a requester's perspective. Knowing the labor process, including the conditions under which it is done and the workers, is key to increasing the visibility of the work [13]. We found that little did the data requesters know about who these annotators were, the working conditions they were under, the hiring requirements they needed to fulfill, and the career aspirations they had. We argue that the impact of knowing the workers and working conditions goes beyond the discussion on the politics of work visibility.

---

[11]Dourish [21] cites Schmidt [52] to make the point that the most influential workplace studies in CSCW have been ones that did not harness themselves to specific design efforts or limit their discussion of implications to then-available design opportunities.

## 6.2 More than just a job, it is a dream

What makes our study particularly interesting is that data annotation is yet another form of full-time employment, especially the fact that it is portrayed as a gateway to the technology and AI industry. People turn to and continue in task-based online works for income [37], a sense of control [8] and flexibility [3, 8]. Gray and Suri [28] argue that since the lack of agreement on the social status and baggage associated with getting data work from the platforms, people pursuing such work made their decisions weighing the costs and benefits according to their situation. We found a clear social status or even prestige attached to working for a data annotation company in our study for our participants. Regardless of what attracted the people to either take part in the platform-facilitated data work or the organized ones through employment, what's common is the lack of long-term stability and career progression. Out of our twenty-five participants, only two could progress substantially in their careers. These job prospects reflect the growing and sobering reality of what is available 1to working-age adults around the globe. As we detailed in our findings, structurally, the annotators are not supported to make meaningful progress in this field or break out of it. Though they might move up annotators to quality analysts quickly, few were able to move beyond senior quality analysts into management. The annotation companies have an organizational structure with an extremely large worker base and a tiny management team. The competition to make it to the top is fierce! There simply aren't enough positions in the chain for the annotators to move forward and upward. The annotators who attempted to break into the more technical world found they didn't have the 'right' work experience. The skillsets and expertise they developed during their annotation tenure did not warranty them a position as an ML engineer or data scientist. In order to obtain the right credentials to transition into the technical field, the annotators need the investment in courses and certificates that serve as the relevant experiences. The irony is that these annotators in our study did come from a technical background with Bachelor's degrees in engineering and technology. Being stuck in annotation and as an annotator means they do not aspire to be anything different or anyone different.

## 6.3 Supporting the aspiration

The expertise of individuals rather than the educational credentials gives one both the employability and a possibility for long-term career trajectory [3, 8, 23]. However, from our research, we found that the expertise the annotators developed during their work seems to have a small application area that does not go beyond the annotation field. Annotation companies made hires based on the educational background while having low previous expertise requirements. The market projection for organized annotation is set to grow as the proliferation of AI systems continues. As the annotation gains increasing popularity, we as researchers need to look at the root cause for the low transferability of annotation skills. We argue that the fundamental skills required for an annotation job are beyond a set of techniques and are currently completely overlooked. We contend that without a clear articulation of the skills, employers will look for proxies such as the educational background to select workers. Whether the 'upskilling' needed for an annotator is additional vocational training aimed at different trades in AI and data at large or simply the knowledge to identify and articulate their skill sets in relevance to other jobs and industries. For instance, the fast learning ability and attention to detail the annotators acquired through extensive domain and bias sensitivity training are qualities important for any job yet overlooked by annotators and their employers. Therefore, training for the job in annotation or industry is of great importance. We ought to ask what is the right or suitable training for annotators to develop the skills needed for work and assist them in pursuing their aspirations, which may or may not be in the annotation industry. The growth in one's career is an aspiration. As discussed above, the pathways upward and forward in annotation are difficult. Though admittedly challenging, there should be a built-in career perspective in the annotation industry. It is also a responsibility that should not be carried alone by the annotation companies. Skill shares, talks, and exposure to the ML/AL systems enabled by the annotator's work can be gateways for an annotator to gain relevant experience transferable to a technical role. Ethical practices around data annotation are yet to break into ethical AI policy and regulatory discussions. This is evident in even the very recent discussions in China and the US concerning algorithms ethics from legal and regulatory perspectives [7, 53]. On a positive note, fair working condition for platform-based gig workers is part of the CAC regulation [7] and are soon to be implemented. At a practice level, as the push for data documentation is picked up by ML practitioners, we too see this as an opportunity where the data labor practice can be documented and reviewed [25, 38, 39]. Since the regulatory measure is yet to be implemented and the documentation practice is still new, we don't know the impact these measures have on either the work of annotation or the annotators. Therefore, substantial research could be done to again look at the impact of these measures from the annotator's perspective. In turn, we argue the insights from annotators would inform improvement in both the policies and practices aimed at protecting their rights.

In our work, what we call upon is a re-centering of data workers as situated and agile experts. The extensive domain and bias sensitivity training they must undergo and how they capitalize on existing forms of expertise must be seen and valued. They are fast-adapting, flexible, and responsive actors to the evolving field of AI and ML must be recognized as qualities that should assist them with career development. This paper is more than an attempt to show that annotators are hidden from view or marginalized in innovation and regulatory processes that are needed. We advocate for systematic and structural changes towards the discretion and care given to data work and data annotators through the collaboration between CSCW/HCI researchers, ML practitioners, legal experts, and policymakers.

## 7 CONCLUSION

Our study provides a first holistic view on data annotation as organized employment, including the perspectives from individual annotators, to managers from annotation companies to the requesters procure and consume annotated data. Though relatively new, the third-party annotation firms span across the globe, hire millions of workers, and set to grow in billions. Our study reveals that the methodical work practices and organized procedures of annotation serve the interest of the annotation companies and requesters more than they do the workers. The professionalization of data annotation jobs offers some benefits of full-time employment over the annotation platforms (e.g., access to pension and insurances). Overall with the pathways for growth are broken, workers are still anxious about employment and performance in the organized annotation. The lack of professional development and progression intensified control through organizational structure, unpaid and much-expected overwork paints a grim picture working in organized annotation employments. We discuss implications for the organization, operations, and data set tasks in working with annotators and call upon our systematic and structural changes centering on the care for both individual annotators and the aspirations they have.

A limitation of the research is the representativeness of the sample. The annotators who provided the majority of insights in our work were mostly from for-profit annotation firms specialized data for autonomous vehicle development. Therefore, annotators from other domain areas and organizations might have different experiences, for instance, the ones who work for impact sourcing annotation firms that specialize in medical image annotation. Additionally, the study was conducted during the COVID pandemic when observation of how their job unfolds in their work settings was impossible. We relied on video call interviews. We were not able to focus on the actual task processes that took place during annotation. Hence we were not able to explore the impact of the task processes on the workers.

## ACKNOWLEDGMENTS
To all the participants who talked to us during a particularly challenging and confusing time. Without the kindness and generosity of these 'strangers', there shall be no research, in particular these qualitative ones.